\documentclass{PoS}

\title{X-ray Reverberation Observational Modelling in Active Galactic Nuclei}

\ShortTitle{X-ray reverberation observational modelling}

\author{\speaker{M. D. Caballero$-$Garcia}$^{1}$,M. Dov\v{c}iak$^1$, M. Bursa$^1$, I. Papadakis$^{2,3}$, V. Karas$^1$\\
 \llap{$^1$} Astronomical Institute of the Academy of Sciences, Bo\v{c}n\'{\i} II 1401, CZ-14100 Praha 4, \\
             Czech Republic, E-mail: \email{garcia@asu.cas.cz}                \\
 \llap{$^2$} Department of Physics and Institute of Theoretical and Computational Physics, \\
             University of Crete, GR-71003 Heraklion, Greece \\
 \llap{$^3$} Foundation for Research and Technology - Hellas, IESL, Voutes, GR-7110 Heraklion, \\
             Greece \\
}

\abstract{X-ray reverberation in Active Galactic Nuclei, believed to be the result of the reprocessing of coronal
photons by the underlying accretion disc, has allowed us to probe the properties of the inner-most regions
of the accretion flow and the central black hole. Our current model (KYNREFREV) computes the time-dependent
reflection spectra of the disc as a response to a flash of primary power-law radiation from a point source
corona located on the axis of the black hole accretion disc (lamp-post geometry). Full relativistic effects
are taken into account. The ionization of the disc is set for each radius according to the amount of the
incident primary flux and the density of the accretion disc. We detect wavy residuals around the best-fit
reverberation model time lags at high frequencies. This result suggests that the simple lamp-post geometry
does not fully explain the X-ray source/disc configuration in Active Galactic Nuclei. There has been 
a noticeable progress into the development of codes for extended coronae (Wilkins et~al. 2016, Chainakun \& Young 2017, 
Taylor \& Reynolds, 2018a,b). Indeed, the model from Chainakun \& Young (2017), consisting of two axial point sources illuminating an accretion disc that produce
the reverberation lags is able to reproduce the observed time-lag versus frequency spectra. The goal of this
paper is to observationally justify the need for an extended corona in order to provide (in the near future) with a mathematical 
formulation of a model for an extended corona in its simplest form. }

\FullConference{Accretion Processes in Cosmic Sources --- II - APCS2018\\
		3--8 September 2018\\
		Saint Petersburg, Russian Federation}

\begin{document}

\section{Introduction}

The first theoretical predictions and simulations of the time evolving echoes of fluorescent lines due
to reflection from relativistic accretion discs around Schwarzschild black holes (BHs) come from the 
90s \cite{Stella1990,CampanaandStella1995}. Since then the models were improved to include rotating Kerr BHs
as well as more realistic reflection spectra where ionisation of the disc is considered. Some of
these models compute not only the light curves and dynamical (time evolving) spectra but also the
resulting frequency dependent lags between different energy bands and energy dependent lags at a
given frequency and thus produce results directly comparable with the observational data. Currently
the most advanced modelling is performed by several groups around the 
world (e.g. \cite{Cacketetal2014,wilkins16,chainakun17,Dovciaketal2019}). Each of these 
models is suited for different use. The code by Wilkins et~al. \cite{wilkins16} is suitable for investigations of photon delays considering
extended corona, where the fluctuations inside the corona are studied. We notice that in these models
for X-ray reverberation there is not any physical heating/cooling mechanism assumed for the corona (for simplicity). The 
model from Chainakun \& Young \cite{chainakun17} may be used inside the spectral fitting package XSPEC \cite{arnaudetal1996} with the
possibility to define an extended corona as well, however, large data tables have to be pre-calculated
first to be able to fit for all the parameters these types of models have and these computations are
quite heavy on computational time (huge servers with many processing cores are needed). On
the other side the model developed in our group (Dov{\v c}iak et~al. \cite{Dovciaketal2019}) aims to be suited for fast
fitting inside XSPEC (every run takes a few seconds only in a single portable computer) on the
expense of a simplified lamp-post geometry provided. This XSPEC model is publicly
available\footnote{https://projects.asu.cas.cz/stronggravity/kynreverb}. 

Models of X-ray reverberation from extended coronae are developed from general relativistic ray tracing simulations. Reverberation 
lags between correlated variability in the directly observed continuum emission and that reflected from the accretion disc arise due 
to the additional light travel time between the corona and reflecting disc. X-ray reverberation is detected from an increasing sample 
of Seyfert galaxies and a number of common properties are observed, including a transition from the characteristic reverberation 
signature at high frequencies to a hard lag within the continuum component at low frequencies (see e.g. \cite{caballerogarcia18} and 
references therein).

This paper is the continuation of a previous work \cite{caballerogarcia17}, which focused on preliminary results obtained by using the code
KYNREFREV by Dov{\v c}iak et~al. \cite{Dovciaketal2019}. This model takes into account a treatment of all the general relativistic effects for accretion discs and light bending 
effects in the disc-corona geometry considering the lamp-post geometry. Whilst in previous works \cite{caballerogarcia18,caballerogarcia17} we focused on the determination of the spin of the BH, here
we compare our results in the context of discussion of possible extended coronal geometries. Sec. 1.1 shows a brief description of the model, Sec. 2 provides
the description of models and results obtained by the use of them considering an extended corona geometry, in Sec. 3 we briefly discuss and compare our results with 
them. Finally, in Sec. 4 we summarize the obtained results and present our plans for future developments in order to provide for a mathematical formulation of a model for an extended corona in its
simplest form.

\subsection{The lamp-post model (Dov{\v c}iak et~al. \cite{Dovciaketal2019} {\tt KYNREFREV})}

Our code (KYNREFREV; \cite{Dovciaketal2019}) takes into account a treatment of all the general relativistic
effects for accretion discs and light bending effects in the disc-corona geometry. Time-lags versus frequency
and energy can be produced and fitted, either in XSPEC or using another suitable tool. This
model computes the X-ray emission from an accretion disc in the lamp-post geometry where the
point-like patch of corona (of zero extension) located on the axis above the black hole illuminates a geometrically thin
Keplerian accretion disc in the equatorial plane. The main physical parameters of the model are
the BH mass and angular momentum, the height of the X-ray source and the observer viewing angle.
The re-processing in the disc is modelled by using the REFLIONX tables \cite{RossandFabian2005}. The radial
ionisation profile is computed from the illuminating flux and the disc ionization is parametrised as a radial
power-law. The reflecting disc region (inner and outer edge and azimuthal segment) as well
as the circular obscuring cloud can be set in the model. The power-law hard lag for frequency
dependence is also included directly in the model.

\section{Existing models for an extended corona}

\subsection{The horizontally and vertically extended corona (Wilkins et~al. \cite{wilkins16}); model A)}

This model has been motivated in order to explain the low-frequency hard lags observed in the time-lag spectra together with the reverberation lags at high frequencies with the
same physical model. So far, there was only an emipirical prescription for the description of the hard-lags. Furthermore, this model has the aim to explain the presence of an ubiquitous
minimum in the reverberation time-lag versus energy spectra at 3\,keV observed in Seyfert galaxies. 

\begin{figure}
\centering
\includegraphics[bb=0 0 267 154,width=7.5cm,angle=0,clip]{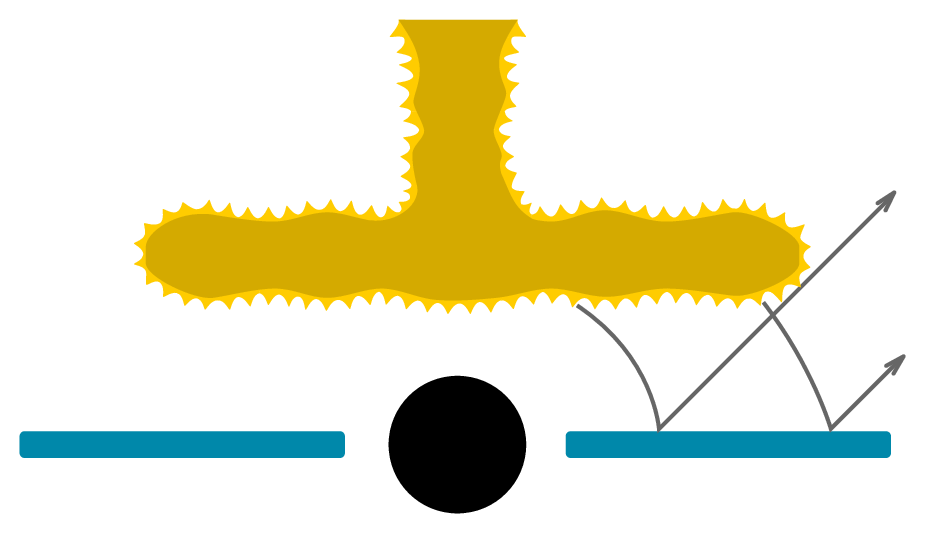}
\includegraphics[bb=0 0 267 113,width=7.5cm,angle=0,clip]{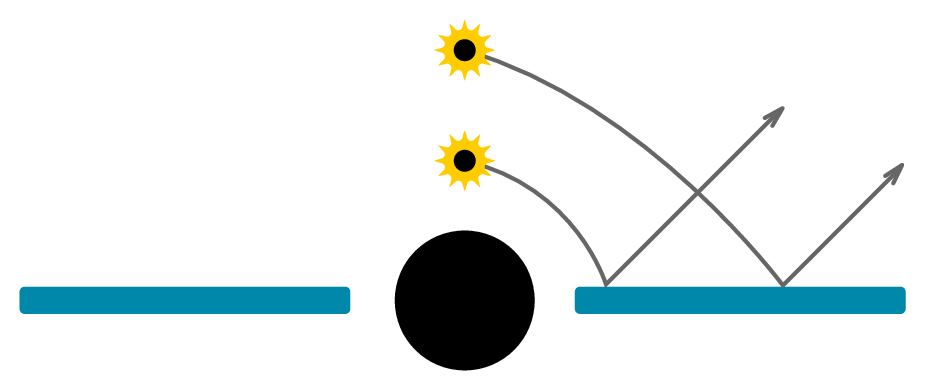}
\includegraphics[bb=0 0 267 125,width=7.5cm,angle=0,clip]{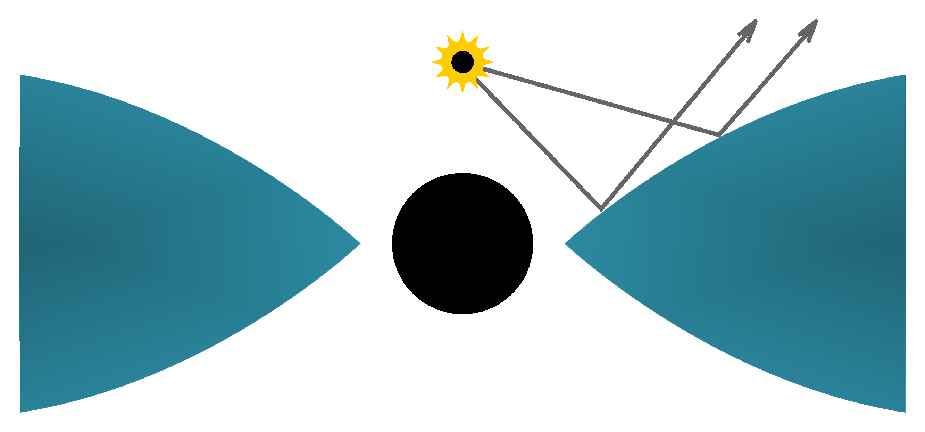}
\caption[]{The sketch of the extended corona model used by Wilkins et~al. (\cite{wilkins16}, model A), the two lamp-post coronae used by Chainakun \& Young (\cite{chainakun17}, model B) and 
the corona with a geometrically thick accretion disc by Taylor \& Reynolds (\cite{taylor18b}, model C).  }
\label{propout.fig}
\end{figure}

Wilkins et~al. \cite{wilkins16} model the reverberation time-lags as produced from extended coronae both over the surface of the disc and vertically. In the model from 
Wilkins et~al. \cite{wilkins16} they refer to horizontal and 
vertical propagations and different physical scenarios are proposed. The corona extends radially at low height over the surface of the disc
but with a bright central region in which fluctuations propagate up the black hole rotation axis (see Fig.~\ref{propout.fig}). We refer the reader to 
Wilkins et~al. \cite{wilkins16} for a proper description of
all the different propagations modes considered by them.

\subsubsection{Comparison with our lamp-post model ({\tt KYNREFREV})}

\begin{figure}
\centering
\includegraphics[bb=0 0 612 792,width=6.3cm,angle=270,clip]{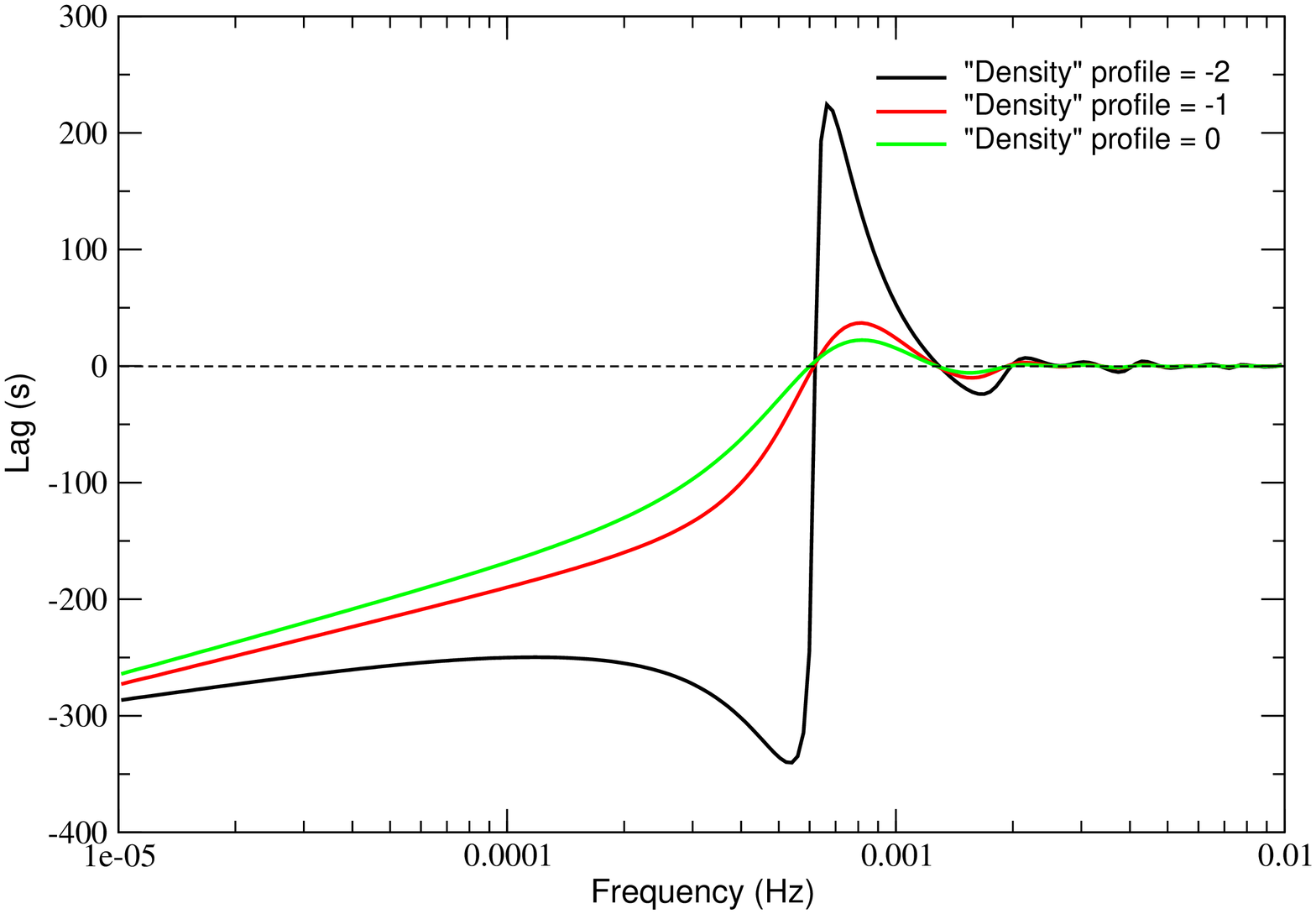}
\includegraphics[bb=0 0 612 792,width=6.3cm,angle=270,clip]{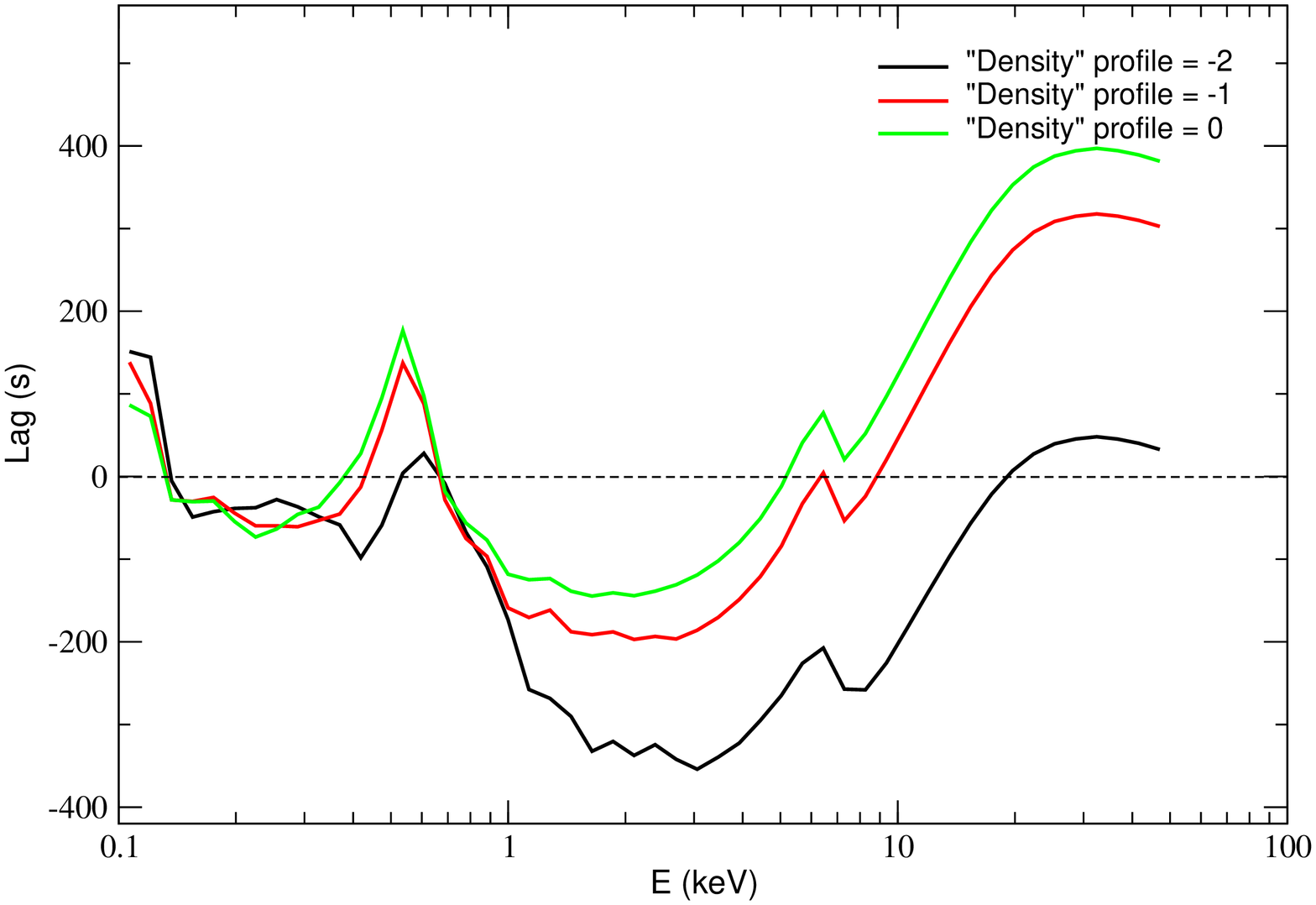}
\includegraphics[bb=0 0 612 792,width=6.3cm,angle=270,clip]{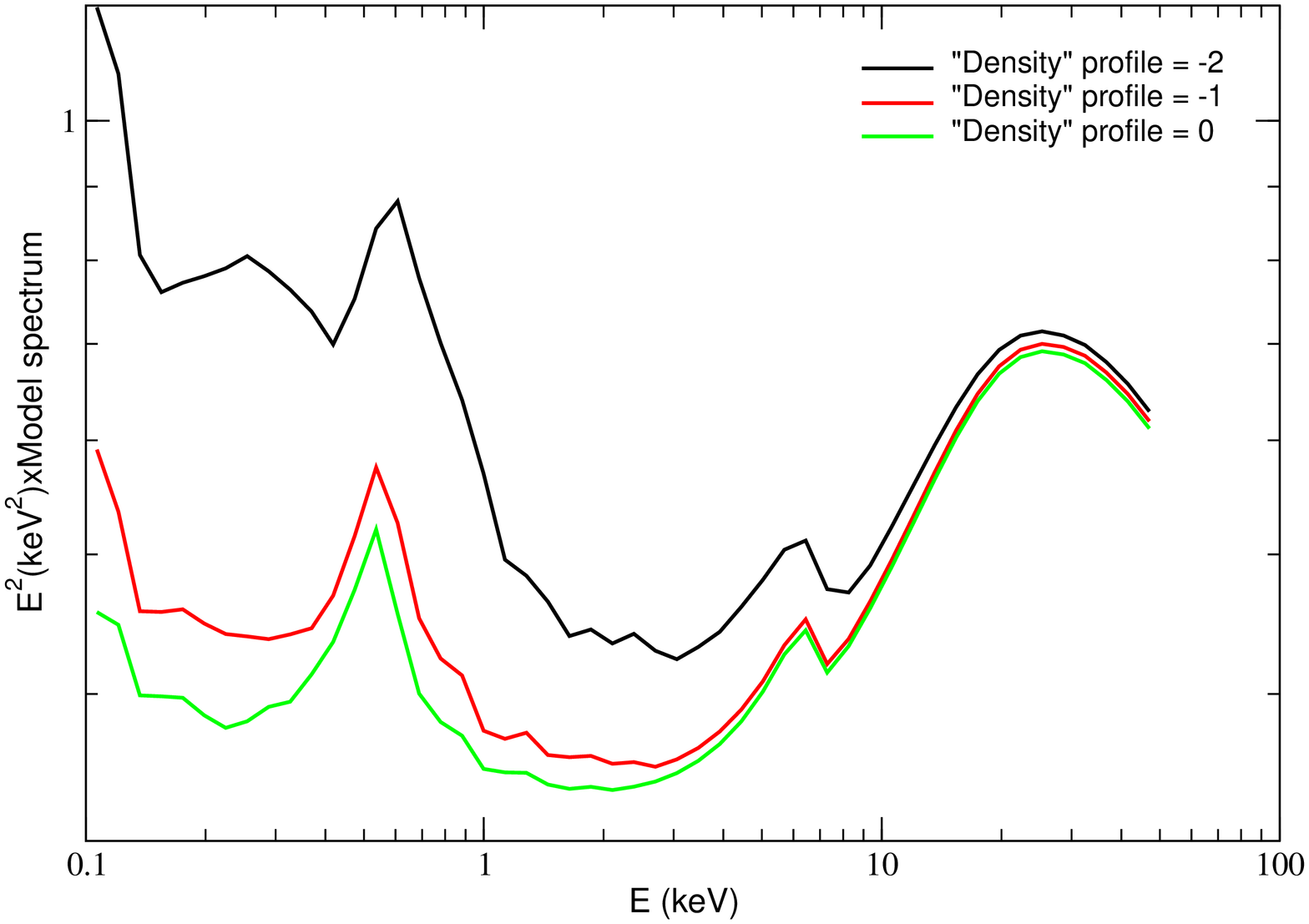}
\caption{Variations with the radial ``density'' profile power-law power index (Top) The dependence of the lag (in seconds) with frequency (Hz) of the emission in the (0.3-0.8\,keV) energy
range with respect to the (1-3\,keV) energy band. (Medium) Lag (in seconds) diluted by primary radiation versus energy (keV) with respect to
the (0.1-10\,keV) energy band at the frequency of $10^{-4}$\,Hz. (Bottom) Average spectrum (i.e. divided by the flare duration) multiplied
by energy in the (0.1-50\,keV) energy range. Different radial power-law ``density'' profile power indexes of -2 (black), -1 (red) and 0 (green)
have been considered, with a ``density'' profile normalization of $10^{15}\,{\rm cm}^{-3}$. The mass of the BH is M=$10^{7}\,{\rm M}_{\odot}$ and
the adimensional spin, inclination of the observer and height of the primary
source are ${\rm a}=1$, ${\theta}=30^{\circ}$ and ${\rm h}=5\,{\rm R}_{\rm g}$, respectively.
}
\label{fig5}
\end{figure}

We have produced time-lags versus frequency and energy (the latter at the frequency of $10^{-4}$\,Hz) and the average
spectrum (divided by the duration of the flare) for a height of the primary of 5${\rm R}_{\rm g}$ and {\it using different values for the power index of the radial
ionization profile of the disc} (parameter {\tt q\_n}=0,-1,-2) \footnote{The one power-law index distribution of density is not realistic. We are working on more realistic assumptions for the density
distribution (work in progress).}. The value for the normalization of the density (parameter {\tt nH0}) has been set to a
canonical value of 10$^{15}\,{\rm cm}^{-3}$ \footnote{Since {\tt REFLIONX} tables are computed only for one value of density, we really cannot change the density, i.e. by
changing the density we mean that we effectively change the ionisation of the disc (from the fact that ${\xi(r,\phi)}{\propto}{\rm F(r,\phi)}/{\rm n(r,\phi)}$ (i.e. note that the dependence of ionization
on $\phi$ are due to the GR effects and may sometimes be omitted hereafter), where ${\xi}$ and $n(r)$ are the
ionization and the density profiles, respectively), i.e. the
ionisation is not only dependent on the radial profile of illumination (given by the lamp-post geometry) but also by radial profile of density.}. The results are 
shown in Fig.~\ref{fig5}. From the plots we can see that when the radial ``density'' distribution
is steeper, the inner disc reflection component dominates. This translates into an increase of the effects due to our relativistic model occurring in the innermost regions
around the BH. Due to the fact that the radial decrease of ionisation is smaller for most radii the contribution from inner disc
part is more ionised and thus there are more reflected photons especially in the soft band. Therefore, the dilution by primary is smaller and
there is an increase in lags (see lag vs. frequency, i.e. top panel, in Fig.~\ref{fig5}). These effects are both the increase of the time lags at the highest 
frequencies (${\gtrsim}10^{-4}$\,Hz) and of the red wing of the relativistic ${\rm Fe}\,{\rm K}_{\alpha}$ line, which is due to emission from the innermost parts of the 
accretion disc. This also explains the lag vs. energy dependence, note that more ionised spectra are steeper (due to larger soft excess).
Visually, this corresponds to the presence
of a pronounced minimum at ${\approx}$3\,keV, occurring when the power index is steep (see their Fig.~21, left panel). We will mention about this fact further in Sec.~\ref{discuss}.

\begin{figure}
\centering
\includegraphics[bb=20 30 612 787,width=6.3cm,angle=270,clip]{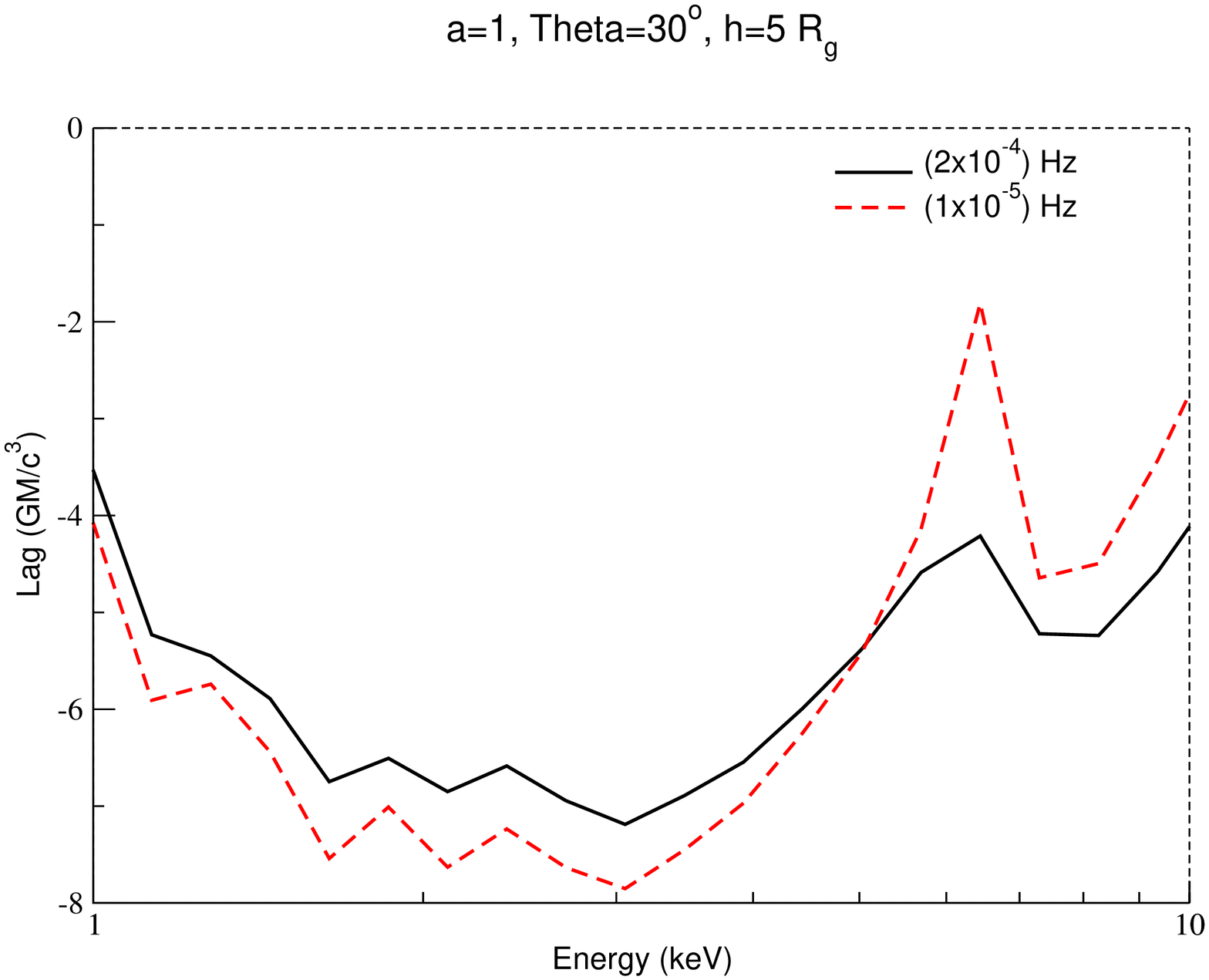}
\includegraphics[bb=20 30 612 787,width=6.3cm,angle=270,clip]{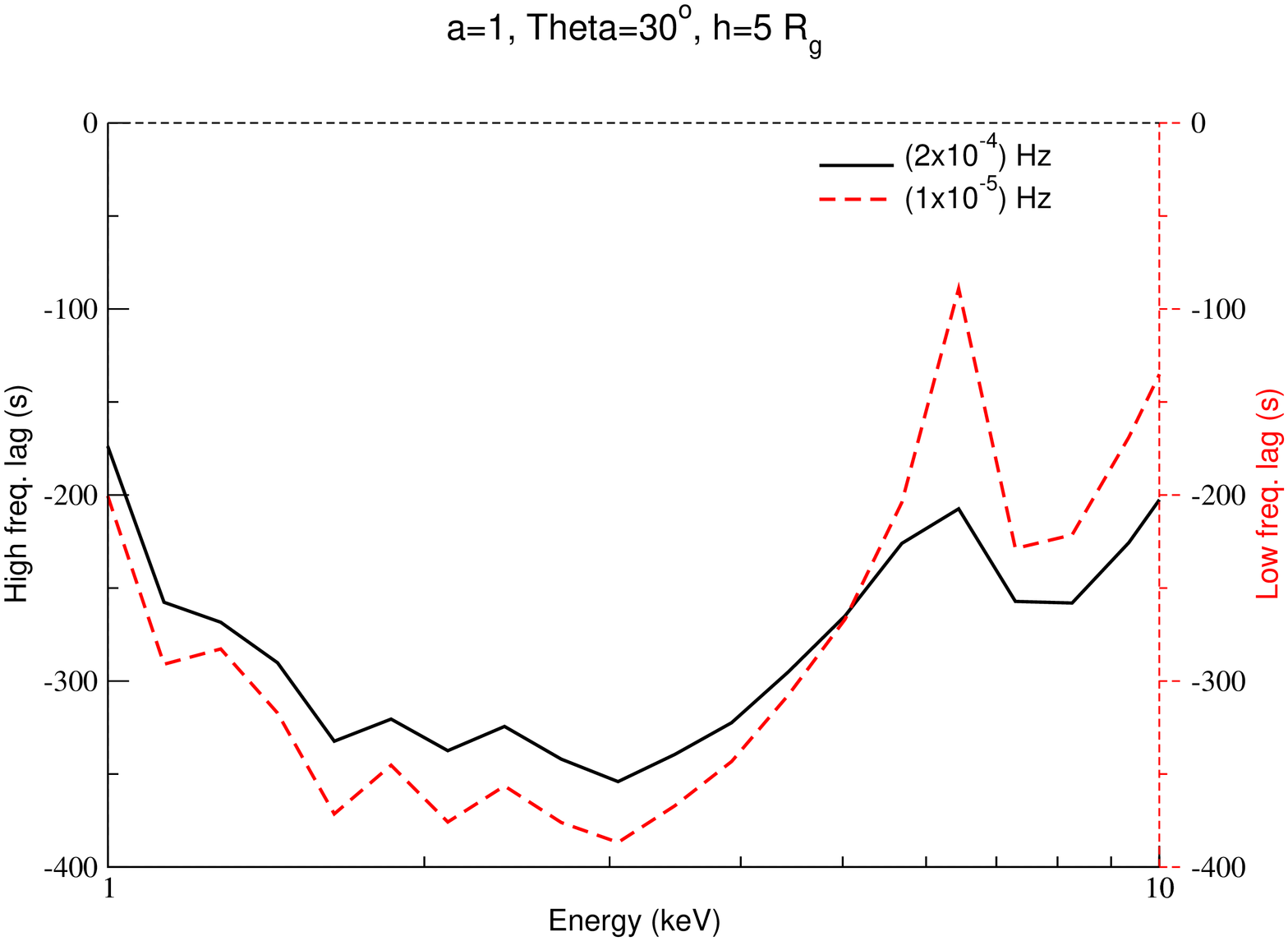}
\caption{Variations of lag-energy plots with frequency. Lag in gravitational units (top) and in seconds (bottom) diluted by primary radiation versus energy (keV) with
respect to the (0.1-10\,keV) energy band at high/low frequencies (solid/dashed lines), at the frequencies
of ($1{\times}10^{-4}$)\,Hz and ($1{\times}10^{-5}$)\,Hz, respectively. A radial power-law density profile power index of -2
has been considered, with a normalization of $10^{15}\,{\rm cm}^{-3}$. The mass of the BH is M=$10^{7}\,{\rm M}_{\odot}$ and
the adimensional spin, inclination of the observer and height of the primary
source are ${\rm a}=1$, ${\theta}=30^{\circ}$ and ${\rm h}=5\,{\rm R}_{\rm g}$, respectively.
}
\label{fig6}
\end{figure}

We also made a further test of obtaining the absolute time lags (where the whole energy band is used as a reference; by setting {\tt E3}=-1) versus energy at low (${\approx}10^{-5}$\,Hz) and
high (${\approx}10^{-4}$\,Hz) frequencies. We used the same values as in Fig.~\ref{fig5} for the spin, inclination angle, height of the primary source, normalization of
the ``density'' and fixed the radial ``density'' profile to -2. The results are shown in Fig.~\ref{fig6}. As it can be seen, the lags are larger and have a narrower peak of the relativistic
${\rm Fe}\,{\rm K}_{\alpha}$ line at low frequencies than at high frequencies. This means that at higher frequencies we are tracking the emission from the matter closer to
the BH, {\it where the relativistic effects are more important}. This broadens the shape of the relativistic ${\rm Fe}\,{\rm K}_{\alpha}$ line, as it can be seen. Further away,
the time lags become longer and the relativistic ${\rm Fe}\,{\rm K}_{\alpha}$ line becomes peaked, because of the more negligible relativistic effects. This is consistent with
the findings of \cite{zoghbi12} (see their Fig.~7).

\subsection{The two blobs extended corona (Chainakun \& Young \cite{chainakun17}); model B)}

Chainakun \& Young (\cite{chainakun17}) present an extended coronal model that consists of two axial point sources illuminating an accretion disc that produce the reverberation lags (Fig.~\ref{propout.fig}). They
assume that these two sources are two lamp-post sources located in the same vertical axis and that both respond independently. Therefore their calculated disc response and the total lags consist of 
a combination of both source and disc responses. They conclude that their analysis of the energy lag spectra of PG~1244+026 shows that the upper
source produces small amounts of reflection and can be interpreted as a relativistic jet.{\it They also suggest that the fluctuations propagating between the two sources of PG~1244+026
are possible but only at near the speed of light}.

\subsubsection{Comparison with our lamp-post model ({\tt KYNREFREV})}

As we mentioned in Caballero-Garcia et~al. \cite{caballerogarcia18} the time-lag versus frequency spectra of a sample of Seyfert Active Galactic Nuclei (AGN) show wavy-residuals which are clearly present in the best-fit residuals, in particular in
the case of Ark~564 and 1H~0707-495 (see the residual plots in Fig.~1 of that paper). These residuals are responsible for the relatively poor statistics of the best-fit models, and suggest that the 
simple lamp-post geometry (which the {\tt KYNREFREV} model assumes) may not be an accurate representation of the X-ray/disc configuration in the inner regions of AGNs.

This may not be surprising since the lamp-post geometry, although widely assumed in the literature, is probably a simplification of the real situation. A more complicated model has been recently proposed 
by Chainakun \& Young \cite{chainakun17}. Their model consists of two co-axial point sources illuminating the accretion disc. It assumes that the variations of two X-ray sources are triggered by the same primary 
variation, but the two sources respond in different ways. The total time-lags are the result of a combination of both the source and disc responses. Their model predictions (see their Fig.~4) appear to be in 
agreement (qualitatively) with the peaky residuals in our Fig.~1 from \cite{caballerogarcia18}. 

\subsection{The geometrically thick disc and lamp-post and/or flaring corona (Taylor \& Reynolds \cite{taylor18b}); model C)}

From the point of view of X-ray spectroscopy Taylor \& Reynolds \cite{taylor18a} found that, at the geometrically thin disc limit their model is consistent with other models taking into account the same approximation, but the spectral models start to diverge significantly 
when $\dot{M} \, > \, 0.1 \, \dot{M}_{\rm Edd}$. With these results, \cite{taylor18a} concluded that accretion disc geometry 
should not be neglected in the detailed modeling of moderate-to-high luminosity AGN reflection spectra, and thus it is reasonable to explore consequences of finite disc thickness in reverberation.

Later Taylor \& Reynolds \cite{taylor18b} proposed a new model to explore the effect that accretion disc geometry has on reverberation signatures, assuming a lamp-post configuration but allowing for a finite disc
scale height (Fig.~\ref{propout.fig}; i.e. geometrically thick disc).

They also show that a flaring corona (approximated by off-axis point-like flares) would not be sufficient to explain the observed high-frequency lags in AGN if the corona has a very low height. {\it This
is consistent with the common view that the corona responsible for the high frequency lag characteristics, such as the Fe K$\alpha$ reverberation, must be physically separated from the reprocessing
material \cite{ReisMiller2013}}.

\subsubsection{Comparison with our lamp-post model ({\tt KYNREFREV})}

As we mentioned in Caballero-Garcia et~al. \cite{caballerogarcia18} the time-lag versus frequency spectra of a sample of Seyfert Active Galactic Nuclei (AGN) 
suggest a compact X-ray source, located close to the BH, at a height of approx. 4 gravitational radii and that this result does not depend either on the BH spin or the disc ionization. Also,
the similarity of the 
wavy-residuals around the best-fit reverberation model time-lags at high frequencies with the model from Chainakun \& Young \cite{chainakun17} is in agreement with the fact that the corona has to be physically separated
from the accretion disc, in agreement with the model of Taylor \& Reynolds \cite{taylor18b}. 

\section{Discussion} \label{discuss}

In this paper we present some interesting scientific outputs obtained by using the {\tt KYNREFREV} model \cite{Dovciaketal2019} which deals with X-ray reverberation effects in AGN and 
can be used in and/or outside XSPEC. This model takes into account all the relativistic effects
of the light surrounding the BH and a realistic prescription of the accretion disc surrounding it under the lamp-post primary source geometry. We undertake the novel approach of considering
a non-constant radial ``density'', illumination and ionization profiles of the accretion disc, which is likely the most realistic situation. This is done by using a variable ionization parameter $\xi(r,\phi)$ as an input for
the {\tt REFLIONX} tables used in our code. Taking this into account we derive some interesting physical consequences. The first one is that for steep radial ``densities'' (${\le}-1,-2$), i.e. for
non-constant ``density'' distributions, the relativistic effects dominate. As a result, the red wing of the relativistic ${\rm Fe}\,{\rm K}_{\alpha}$ line and from the so-called ``soft-excess'' appear broader in the lag-energy spectra
when compared to the non-relativistic case (as is the case of the reflection in the outer regions of the disc, where we also expect longer time-delays versus energy). This produces a shift 
to higher energies of the minimum to 3\,keV in the lag-energy spectra, which coincides with the spectral region where the power-law emission from the primary
source is dominating, hence the relativistic effects are minimal.

The second one is that the innermost regions are more affected by GR effects than the outer regions. This means
that the highest frequency lags (produced closer to the BH) have a broader red relativistic ${\rm Fe}\,{\rm K}_{\alpha}$ line wing than the low frequency ones (produced further away
from the BH), which are peaked. This fact agrees with what it was already pointed out in previous studies \cite{zoghbi12} so we will not discuss it any further.

Both the first and second physical consequences derived by using this model coincide with what it is theoretically expected by taking into account pure GR effects in the most
realistic accretion disc scenario (i.e. with variable ``density'', illumination and ionization radial profile) under the simple lamp-post geometry. More complicated prescriptions of the corona
can be considered, as for example the use of extended coronae. This work has been done already \cite{wilkins16}. They have developed an X-ray reverberation code taking into
account all the effects due to GR under the approximation of a constant ``density'' profile. They find that, in order to produce the pronounced minimum at 3\,keV in the lag-spectra, which
has been observed in a few AGN, an horizontally and {\it vertically} extended corona is needed. We emphasize here that only a vertically extended corona, with propagations
travelling at $v{\approx}$1\% of the speed of light, under the assumption of constant properties of the disc, can reproduce the observed reverberation lags (see their Fig.~21). {\it We notice that in
the case of vertical propagation at speed of light ($v=c$), the model by Wilkins et~al. \cite{wilkins16} is not able to reproduce the observed minimum at 3\,keV (see their Fig.~21, left panel), that we 
naturally reproduce with our lamp-post (and ionized disc) model}.  

\section{Summary and conclusions}

The results of the comparison between our model and these ones can be summarised as follows:

\begin{itemize}
\item{The time-lags observed from a sample of Seyfert 1 AGN (Caballero-Garcia et~al. \cite{caballerogarcia18}) show residuals compatible with the two-blobs extended corona model (Chainakun \& Young \cite{chainakun17}). }
\item{In the lamp-post approximation and taking into account realistic accretion rates and/or disc thickness of the accretion disc show that very low coronal heights ($h{\le}3\,{\rm r}_{\rm g}$)
are not possible(observed) (Taylor \& Reynolds \cite{taylor18b}), in agreement also with our findings from a study of the data of a sample of Seyfert 1 AGN (Caballero-Garcia et~al. \cite{caballerogarcia18}). }
\item{From all these studies one can say that: there are strong indications from the reverberation X-ray mapping that most of the irradiating photons must come from a source(corona) that is physically separated from the reprocessing
material, such as a (zero-extent) lamp-post or an extended jet.  }
\end{itemize}

From the total phase of the Fourier transform of the total flux (i.e. both the primary and the reflection continuum are taken into account) one can calculate the time-lag of the signal. This is computed
from the total phase at energy bin E with respect to the total phase at some reference energy bin (see \cite{caballerogarcia17}). This can be done from each blob of the corona as emitting in the form of 
a power-law (for simplicity). We note that any further developments of this formulation will come in future work.

\acknowledgments

\noindent MCG, MD and MB acknowledge support from the Grant Agency of the Czech Republic within the project n$^{\circ}$\,18-00533S. MCG acknowledges funding from ESA through a partnership with IAA-CSIC (Spain).

\end{document}